\documentclass[aps,pra,superscriptaddress,twocolumn]{revtex4-1}
\usepackage[]{graphicx}
\usepackage[]{textcomp}
\usepackage[]{xspace}
\usepackage{dcolumn}
\usepackage{epstopdf}

\begin{document}
\title{Coherence area profiling in multi-spatial-mode squeezed states}
\date{\today}
\author{B. J. Lawrie$^{1*}$ \& R. C. Pooser$^{1}$ \&  N. Otterstrom$^{1,2}$\\ $^1$Quantum Information Science Group, Computational Science and Engineering Division, Oak Ridge National Laboratory, Oak Ridge, TN\\$^2$ Department of Physics \& Astronomy, Brigham Young University, Provo, Utah 84602, USA.\\$^*$lawriebj@ornl.gov}

\begin{abstract}
The presence of multiple bipartite entangled modes in squeezed states generated by four wave mixing in atomic vapors enables ultra-trace sensing, imaging, and metrology applications that are impossible to achieve with single-spatial-mode squeezed states. For Gaussian seed beams, the spatial distribution of bipartite entangled modes, or coherence areas, across each beam is largely dependent on the spatial modes present in the pump beam, but it has proven difficult to map the distribution of these coherence areas in frequency and space. We demonstrate an accessible method to map the distribution of the coherence areas within these twin beams. We also show that the pump shape can impart different noise properties to each coherence area, and that it is possible to select and detect coherence areas with optimal squeezing with this approach.\end{abstract}

\maketitle



Four-wave-mixing~\cite{McCormick2007OL} in alkali vapors can amplify many spatial modes, or coherence areas~\cite{Jedrkiewicz2004}, resulting in a two-mode squeezed state for each coherence area pair across the resulting twin beams~\cite{Boyer2008,Lawrie2013,corzo2011multi,clark2012}. 
The total number of coherence areas is dependent on the angular acceptance bandwidth of the four-wave-mixing geometry, and the distribution of coherence areas is strongly dependent on the pump and probe intensities, beam profiles, and waist positions, along with the pump and probe frequency detuning~\cite{Boyer2008,boyer2008generation,Lawrie2013}. Because these states exhibit noise below the photon shot noise limit (SNL) in multiple coherence areas, they enable quantum enhanced imaging, nonlinear interferometry, and quantum plasmonic sensing and imaging platforms~\cite{Lassen2007,Kolobov2000,Lawrie2013,lawrie2013prl,clark2012,hudelist2014quantum,pooser2014ultrasensitive,treps2002surpassing,treps2003,treps_multipixel}. However, until now there has been no clear visualization of those coherence areas within twin beams that did not consist of well separated spatial modes created by transferring a high contrast image to the seed beam~\cite{Boyer2008,Lawrie2013}. For applications such as optical beam deflection, where specific spatial modes are necessary to optimize the signal to noise ratio (SNR) in beam position measurements on split photodiodes~\cite{treps2002surpassing,pooser2014ultrasensitive}, these highly structured squeezed states are generally sub-optimal. We demonstrate an accessible method to characterize the distribution of quantum correlations shared between pairs of coherence areas. With this approach, controlling the spatial profile and frequency detuning of the pump field enables precise control over the distribution of coherence areas.  For specific spatial filter configurations and pump detunings, we show that the excess noise in some spatial modes can be filtered, resulting in increased quantum correlations in attenuated beams.

\begin{figure}[h!]
\centerline{\includegraphics[width=3.65in]{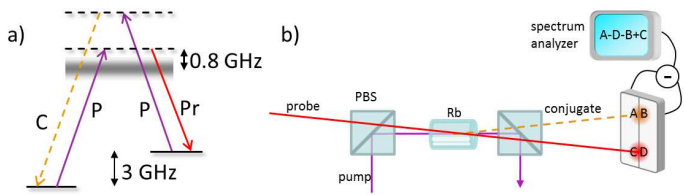}}
\caption{(a) Energy level diagram illustrating double $\Lambda$ system in ${}^{85}$Rb at the D1 line (795 nm). (b) Schematic of squeezing measurement performed on split photodiode with weak probe beam of 50 $\mu $W offset by 3.045 GHz from a 100 mW pump beam.}
\label{fig:figure1}
\end{figure}

The split photodiodes that are used for MEMS beam displacement measurements enable a simple technique for mapping coherence areas.  As shown in Fig. \ref{fig:figure1}, it is possible to measure quantum noise reduction when the probe and conjugate twin beams generated by four-wave-mixing in hot ${}^{85}$Rb vapor are equally divided on a high quantum efficiency split photodiode \cite{pooser2014ultrasensitive}.  If the coherence areas in each beam were well isolated on each side of the split photodiode, then the measured noise would be identical to that measured when the probe and conjugate were detected by separate channels on a balanced photodiode, or: 
\begin{equation}
\langle \Delta N^2 \rangle = \frac{1}{\eta (2G-1)}
 \label{eq:x1}
\end{equation}
where $\eta$ is the combined detection efficiency and G is the four wave mixing gain. However, the coherence areas will generally not be perfectly split between two halves.  As a result, coherence areas distributed over both detector elements contribute fractions of shot noise units to the combined measurement noise \cite{treps_multipixel}. For twin beams consisting of multiple coherence areas with equal gain, the noise is given by:

\begin{equation}
\langle \Delta N^2 \rangle = {\left[ \frac{1}{P_{0}} \left( P_{sw} \eta_{d} \left( 2G-1 \right)+\displaystyle\sum_{i=1}^{N} P_{i}\eta_{i}(2G-1)\right) \right]}^{-1},
 \label{eq:x2}
\end{equation}
where $P_{sw}/P_{0}$ is the fraction of optical power in coherence areas isolated on one channel of the split detector,  $P_i$ is the power partially incident on a detector channel from the $i^{th}$ mode such that $\sum_{i=1}^{N} P_{i} =P_{0}-P_{sw}$, $\eta_{d}$ is the combined detector efficiency, $\eta_{i}$ is the detection efficiency of the $i^{th}$ mode (including attenuation resulting from being split by the detector), and $N$ is the total number of modes split across the detector. 

A misalignment which results in a splitting of coherence areas across the two detector halves adds both fractions of shot noise units and excess noise associated with the anticorrelated coherence area pairs. If the probe and conjugate coherence areas are split in the same way (as mirror images of one another in their image planes), then the excess noise is subtracted within the detector's common mode rejection limits, and Eq.~\ref{eq:x2} applies. In this case, only fractions of shot noise units are added to the noise floor, resulting in a linear dependence of the quantum noise reduction on the fraction of misaligned coherence areas \cite{fabre2000quantum}.

In the limit that a single coherence area in the probe and conjugate channels is evenly split, the noise in Eq.~\ref{eq:x2} reduces to twice Eq.~\ref{eq:x1}, or a 3 dB increase in noise for a perfectly split single coherence area, indicating that in spatially filtered measurements (such as imaging or beam displacement), a multimode beam with a large number of coherence areas always outperforms single mode beams.

The coherence areas in the probe and conjugate beams demonstrate inversion symmetry around the center of the pump beam in their respective image planes as a result of conservation of momentum in the four-wave-mixing process. Thus, by sweeping the probe and conjugate beams across the split detector from opposite sides in their respective image planes, the measured quantum noise reduction will be maximized when coherence areas are largely isolated on an individual channel, and minimized when coherence areas are split across both channels.  When the probe and conjugate fields are in their respective image planes at the split detector, it is straightforward to demonstrate optimal alignment of coherence areas on the appropriate detectors, but if either beam is not in its image plane, the distortion of the coherence area distribution relative to the other channel will result in excess noise. Other authors have previously rastered razor blades symmetrically and anti-symmetrically across the probe and conjugate in order to demonstrate the presence of multiple spatial modes \cite{Boyer2008,corzo2011multi,Marino2008}, but those approaches did not yield any information about the distribution of coherence areas.  

Creating a pseudo-split detector by using D-mirrors to select the modes A,B,C, and D illustrated in Fig.~\ref{fig:figure1} before mixing modes A and C and B and D on each channel of a balanced photodiode allows an accurate reproduction of a split detection measurement and also allows the straightforward measurement of correlations between A and D or B and C by simply blocking the necessary channels.  The combination of these noise measurements enables accessible coherence area mapping in quantum correlated twin beams.
 
\begin{figure}[h]
\centerline{\includegraphics[width=9cm]{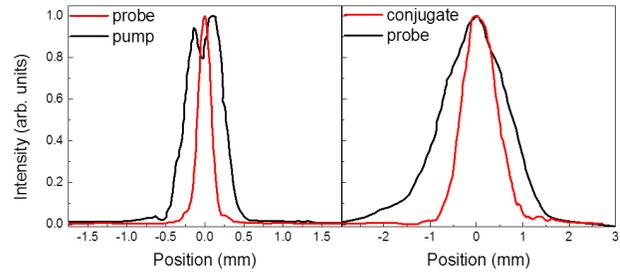}}
\caption{(a) Horizontal probe and pump beam profiles at their respective waists, and (b) the probe and conjugate beam profiles in their respective image planes after four-wave-mixing.}
\label{fig:figure2}
\end{figure}

The squeezed state used in this experiment was generated via four wave mixing with a weak seed probe beam of 50 $\mu$W offset by 3.045 GHz from a 100 mW pump beam in a one-inch long ${}^{85}$Rb vapor cell whose stem was held at a temperature of 80~$\pm 1^\circ$~C. The two beams intersected at their waists in the vapor cell at an angle of 7 mrad. The probe exhibited a Gaussian beam profile at its waist, while the waist of the pump beam exhibited a bimodal structure as shown in Fig.~\ref{fig:figure2}a. The image planes of the probe and conjugate fields were determined by measuring the quantum noise reduction present between the two channels as D mirrors were used to attenuate 50\% of each beam.  When either D mirror was far from its beam's image plane, the distorted coherence areas yielded significant anti-squeezing between channels A and D or B and C.  On the other hand, when each D-mirror was near its beam's image plane, quantum correlations were present and increased squeezing was measured. The probe and conjugate image planes could thus be identified by optimizing the quantum noise reduction between channels A and D as a function of D-mirror position along the propagation axis. Under the above conditions, blueshifting the pump frequency 1.45 GHz from the F=2 to excited state transition resulted in 4.5 dB of quantum noise reduction (corresponding to a gain of 12.6 with 85\% transmission on the probe channel).  

The large gain required for optimal squeezing in our experimental configuration resulted in a strong Kerr lensing effect that set the probe and conjugate image planes close to the vapor cell, making them difficult to isolate from the scattered pump field. Red-shifting the pump reduced the Kerr lensing effect at the expense of reduced squeezing. In particular, red-shifting the pump by 250~MHz resulted in 1.4~$\pm 0.05$ dB of squeezing with probe and conjugate image planes located 94~cm and 32~cm from the center of the vapor cell respectively. At these image planes, the probe and conjugate beams demonstrated Gaussian profiles with 1.6 mm and 0.8 mm spot sizes (full-width-half-max) as shown in Fig.~\ref{fig:figure2}b. Other configurations in 4WM can result in optimal squeezing with separable beams~\cite{Boyer2008,Lawrie2013,corzo2011multi}, but our approach would work for any multispatial mode squeezed states with accessible image planes~\cite{treps2002surpassing,treps2003}.

\begin{figure}[ht!]
\centerline{\includegraphics[width=3.5in]{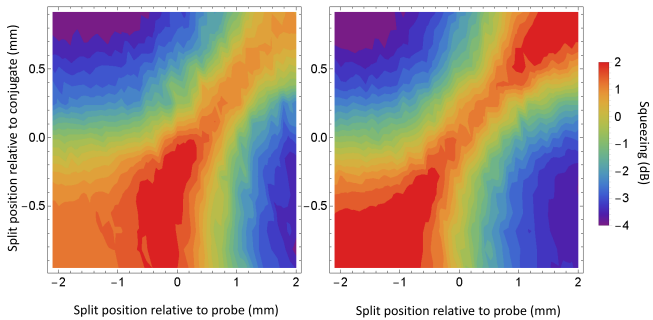}}
\caption{Horizontal noise distribution across probe and conjugate fields as a function of probe and conjugate D-mirror positions. Left: B and C modes are blocked while A and D are collected on a balanced detector.  Right: the split detector configuration shows the spatial mode structure across the two beams.}
\label{fig:figure3}
\end{figure}

Figure \ref{fig:figure3} illustrates the spatially resolved noise distribution across the probe and conjugate beams for both split and knife edge configurations.  The conjugate mirror was rastered in 15 steps across the conjugate beam, while the probe mirror was rastered in 40 steps across the probe beam for each conjugate mirror position, resulting in 600 total noise measurements.

\begin{figure}[htb]
\centerline{\includegraphics[width=3.25in]{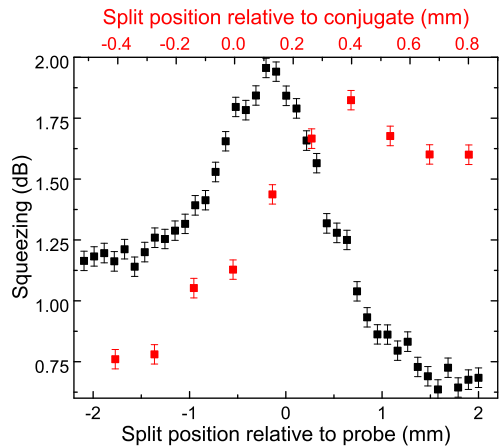}}
\caption{Horizontal noise distribution across probe and conjugate fields as measured on channels A and D.}
\label{fig:figure4}
\end{figure}

As has been shown previously~\cite{Boyer2008,Lawrie2013,corzo2011multi}, the presence of anti-squeezing everywhere in Fig.~\ref{fig:figure3} except where equal fractions of each channel were blocked symmetrically about the pump beam is evidence of the multi-spatial-mode nature of the quantum correlations in this two mode squeezed state. The asymmetric structure present in the right hand side of Fig.~\ref{fig:figure3} shows the pump-dependent distribution of spatial modes, while the left hand side shows the isolation of a highly correlated coherence area off center in each beam. The angular bandwidth of the four wave mixing process in our setup supports approximately 70 modes, estimated using a mode counting technique previously reported \cite{boyer2008generation,Lawrie2013}, but the probe and conjugate fields here only overlap with a small fraction of this bandwidth.  The measured beam diameters in the far field suggest that 3-4 coherence areas are present within the full width half max of the probe and conjugate fields. By plotting the optimal quantum noise reduction for each step of the probe D-mirror and for each step of the conjugate D-mirror as shown in Fig.~\ref{fig:figure4}, it is possible to resolve the structure of the coherence areas within each beam without using high contrast images to isolate them~\cite{Boyer2008}.

While the four wave mixing process resulted in no significant change to the probe beam profile in Fig. \ref{fig:figure2}, Fig.~\ref{fig:figure4} illustrates a clear spatial structure in the observed quantum noise reduction. Squeezed states are highly sensitive to optical attenuation, but the squeezing shown in Fig.~\ref{fig:figure4} and Fig.~\ref{fig:figure3} increases as the left-hand side of the probe and the right-hand side of the conjugate in channels B and C are blocked.  The squeezing is nearly maximized when half of the probe field is blocked, indicating an asymmetric, split structure in the coherence area distribution.

Because a similar fraction must be removed from the conjugate beam in order to increase the squeezing, the effects observed in Fig.~\ref{fig:figure4} cannot be attributed to misalignment between the pump and probe in the vapor cell. Further, when the probe beam was blocked before the vapor cell, scattered pump light resulted in less than 1 dB of noise above the electronics floor.  It is therefore clear from Fig.~\ref{fig:figure4} that there exist at least two coherence areas in this squeezed state.  One, on the left of the probe and right of the conjugate, demonstrates minimal quantum correlations with less than 1 dB of quantum noise reduction.  The other, on the right of the probe and left of the conjugate demonstrates stronger quantum correlations with roughly 2 dB of quantum noise reduction. Closer to resonance (250~MHz to the red), aperturing the probe and conjugate beams did not yield increased quantum noise reduction in this manner, indicating that all coherence areas exhibit comparable quantum noise reduction when the pump is closer to resonance. Thus, the pump shape and detuning both play a role in the arrangement of quantum correlations across coherence areas in the far field.

\begin{figure}[t!]
\centerline{\includegraphics[width=3.25in]{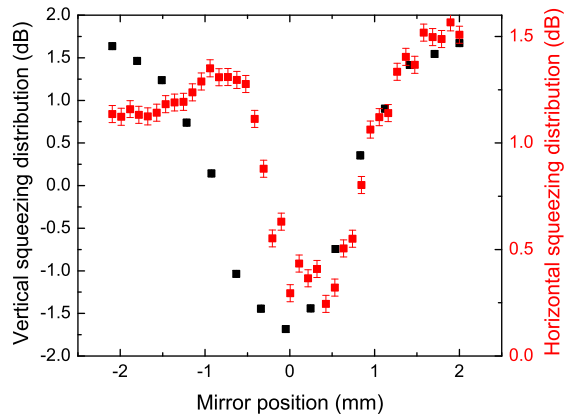}}
\caption{Vertically and horizontally resolved squeezing distribution separately measured at the same pump frequency provide evidence for a single coherence area when swept vertically, but for multiple coherence areas when swept horizontally. All four modes are collected on a balanced detector. }
\label{fig:figure5}
\end{figure}

The approach taken in Fig.~\ref{fig:figure4} is comparable to demonstrations that have previously used spatial filters to demonstrate multi-spatial mode characteristics though the evidence of coherence area structure in a Gaussian beam has never been previously demonstrated.  Unfortunately, optical attenuation ultimately reduces the signal sufficiently that it is difficult to profile the quantum correlations in the entire beam in this way, and the presence of a very weakly correlated coherence area means that a plot of the noise distribution on channels B and C is dominated by excess noise as the quantum correlated portion of the beam is attenuated.  However, by combining all four modes on channels A, B, C, and D, it is possible to develop a more complete profile of the coherence area distribution. Figure~\ref{fig:figure5} illustrates a map of the noise distribution on all four channels when D mirrors are swept vertically and horizontally across the probe and conjugate fields. No structure is evident in the vertical profile, and the measured noise increases by 3.3 dB when the conjugate and probe are evenly split between A and B and C and D respectively.  This suggests that the coherence areas in the probe and conjugate are distributed horizontally across the beams (as the twin beams appear to consist of single coherence areas when profiled vertically). The fact that slightly greater than 3 dB of increased noise was observed is consistent with a small amount of light scattering from the edge of each D mirror.

On the other hand, compared with the structureless vertical scan, the horizontal scan illustrates the initial increase in squeezing previously seen in Fig.~\ref{fig:figure4}, but also a small local maximum in the squeezing near the center of the beam. This local maximum is difficult to resolve in Fig.~\ref{fig:figure3} because of the larger range of measured noise, but it is reproducible and falls outside of the error bars of our noise measurements, suggesting that at least three coherence areas are aligned horizontally in the probe and conjugate fields.  The previously discussed coherence area on the left of the probe exhibits minimal squeezing between probe and conjugate, while evidence exists for an additional two squeezed coherence areas split horizontally at 0.2 mm on Fig.~\ref{fig:figure5}.  When each of those coherence areas is located primarily in only one channel of our emulated split detector, a slight decrease in quantum noise is recorded. 

The emulated split detector approach to profiling the distribution of quantum correlations in multi-spatial-mode squeezed states outlined in this manuscript provides a valuable tool that is critical to many quantum metrology and sensing applications. The horizontal coherence area distribution illustrated in Fig.~\ref{fig:figure5} was not observed for a Gaussian pump beam profile or for pump fields closer to resonance with the D1 transition, indicating the importance of pump beam shaping to tailored coherence areas. The approach to coherence area mapping that we have described here can be utilized in concert with spatial and frequency control over the pump field in order to engineer desired distributions of quantum correlations in two mode squeezed states.

This work was performed at Oak Ridge National Laboratory, operated by UT-Battelle for the U.S. Department of energy under contract no. DE-AC05-00OR22725, and was supported in part by the U.S. Department of Energy, Office of Science, Office of Workforce Development for Teachers and Scientists (WDTS) under the SULI program. B.~L.~and R.~C.~P~ acknowledge support from the lab directed research and development program.

\end{document}